\newcommand{\slaninafiginline}[1]{#1}
\def\slaninafigsize{ voffset=-420 hoffset=-60 
        hscale=80 vscale=80 }
\def\slaninafigspace{50mm} 
\def\slaninafigsizeh{ voffset=-420 hoffset=-60 
        hscale=80 vscale=80 }
\def\slaninafigspaceh{80mm} 
\documentstyle[prl,aps]{revtex}
\begin{document}
\draft

\twocolumn[\hsize\textwidth\columnwidth\hsize\csname@twocolumnfalse\endcsname

\title{On the possibility of optimal investment}

\author{Franti\v{s}ek Slanina}
\address{Institute of Physics,
	Academy of Sciences of the Czech Republic
	Na~Slovance~2, CZ-18221~Praha,
	Czech Republic\\
	and Center for Theoretical Study
	Jilsk\'a~1,CZ-11000~Praha, Czech
	Republic\\
	e-mail: slanina@fzu.cz}

\maketitle
\begin{abstract}
We analyze the theory of optimal 
investment in risky assets, developed recently by Marsili, Maslov and Zhang
{[Physica A 253 (1998) 403]}.
When the real 
data are used instead of abstract stochastic process, it appears that
a non-trivial investment strategy   
is rarely possible. We show that non-zero transaction costs make the
applicability  of the method even more difficult. We generalize the
method in order to take into account possible correlations in the
asset price.
 \end{abstract}

\twocolumn]

\section{Introduction}
Non-equilibrium statistical mechanics, especially the theory of
stochastic processes, finds recently wide applicability in economics. 
First area, intensively studied in the last several years, is the
phenomenology of the signal (price, production, and other economic variables)
measured on the economics system 
\cite{mantegna_91,am_bu_ha_97,li_ci_me_pe_sta_97,go_me_am_sta_98,ga_ca_ma_zha_97,va_bo_mi_au_98,so_jo_97,sornette_98d}.
Scaling concepts proved to be a very useful tool for such analysis.

Second area concerns optimization. In the competitive
economics, agents should maximize their survival probability by
balancing several requirements, often mutually exclusive, like profit
and risk
\cite{bou_so_94,bou_po_97,ga_bou_po_98,bouchaud_98,cont_98}. Third
area comprises creation of models which should grasp particular
features of the behavior of real economics, like price
fluctuations
\cite{ca_ma_zha_97,ba_pa_shu_97,cha_zha_97,sa_ta_98,co_bou_97,bou_co_98}.

We focus here on an aspect of optimization, discussed recently by
Marsili, Maslov and Zhang \cite{ma_mas_zha_98}.
In a simplified version of the economy, there are two possibilities
where to put a cash: to buy either a risky asset (we shall call it
stock, but it can be any kind of asset)
or a riskless asset (deposit in a bank).
In the  latter case we are sure to gain each
year a fixed amount, according to the  interest rate. On the contrary,
putting the money entirely to the stock is risky,  but the gain may be
larger (sometimes quite substantially). We may imagine,  that
increasing our degrees of freedom by putting a specified portion  of
our capital into the stock and the rest to the bank may lead to
increased growth of our wealth. This way was first studied by
Kelly and followers \cite{kelly_56,thorp_71} and intensively
re-investigated recently
\cite{ma_mas_zha_98,mas_zha_98,mas_zha_98a,ba_pa_se_vu_98,au_ba_ha_se_vu_98,serva_98}. 

The point of the Kelly's approach is, that if we suppose that the
stock price performs a multiplicative process
\cite{deutsch_94,sornette_98c,so_co_97,sornette_97a}, the quantity to 
maximize is not the average value of the capital, but the typical
value, which may be substantially different, if the process is
dominated by rare big events.
It was found that given
the probability 
distribution of the stock price changes, there is a unique optimal
value of the fraction  of the investor's capital put into the
stock.

The purpose of the present work  is to investigate the
practical applicability of the strategy suggested in
\cite{ma_mas_zha_98,mas_zha_98}. Let us first briefly summarize this
approach. 
We suppose that the price $p_t$ of the stock changes from time $t$ to
$t+1$ 
according to a simple  multiplicative process
\begin{equation}
p_{t+1} = p_t \, {\rm e}^{\eta_t}
\end{equation}
where $\eta_t$ for different $t$ are independent and equally
distributed random variables. The angle brackets $<>$ will denote
average over these variables.

We denote $W_t$ the total capital  of the investor at the moment $t$. The
fraction $r$ of the capital is stored in stock and the rest is
deposited in a bank. We will call the number $r$  investment
ratio. The 
interest rate provided by the bank  is supposed to be fixed and equal to $\rho$
per one time 
unit. The strategy of the investor consists in keeping the investment
ratio constant. It means, that he/she sells certain amount of stock
every time the stock price rose and sell when the price went down.  

If we suppose that the investor updates its portfolio (i.~e. buys or
sells some stock in order to keep the investment ratio 
constant)  at each time step, then starting from the capital $W_0$,
after $N$ time steps the investor owns  
\begin{equation}
W_N = \prod_{t = 0}^{N-1} (1+\rho + r ({\rm e}^{\eta_t} - 1 - \rho))
W_0 \;\; .
\end{equation}

The formula can be simply generalized to the situation when there is a non-zero
transaction cost equal to $\gamma$ (see also \cite{serva_98}) and the
update of the portfolio is done each  
$M$ time steps. We assume for simplicity that $N$ is
a multiple of $M$. 
\begin{equation}
W_N = \prod_{t = 0}^{N/M-1} { (1+\rho)^M + r ({\rm e}^{\bar{\eta}_{Mt}}
   (1 + G) - (1 + \rho)^M)
    \over 1 + r G} W_0
\end{equation}
where we denoted
$
\bar{\eta}_{Mt} = \sum_{i=Mt}^{Mt+M-1} \eta_i \; 
$
and 
$
G=\gamma \;{\rm sign}(M\ln (1 + \rho) - \bar{\eta}_{Mt})\;\; .
$

We can see that like the stock price itself, the capital performs a 
multiplicative process. 
\begin{equation}
W_{t+1} = e_t(r) W_t
\end{equation}
where the random variables $e_t(r)$ depend on the investment ratio as a 
parameter.

For $N$ sufficiently large the typical growth of the capital
$(W_{t+1}/W_t)_{typical}$ is not equal to the mean $<e(r)>$ as one would 
naively expect, but is given by the median \cite{ma_mas_zha_98}, which
in this case gives  
\begin{equation}
g(r) = \log((W_{t+1}/W_t)_{typical}) = <\log e(r)>\; .
\label{eq:gain}
\end{equation}

Therefore we look for the maximum of $g$ as a function of $r$, which
in the simplest case without transaction costs leads to the equation
\begin{equation}
<{ {\rm e}^\eta - 1 - \rho    \over 
1 + \rho + r_{\rm opt} ({\rm e}^\eta - 1 - \rho )}> = 0 \; .
\label{eq:forr}
\end{equation}
for the optimum strategy $r_{\rm opt}$. If the solution falls outside the interval $[0,1]$,
one of the boundary points is the true optimum, based on the following
conditions. 
If $g^\prime(0) < 0$ the optimum is $r_{\rm opt} = 0$.
If $g^\prime(1) > 0$ the optimum is $r_{\rm opt} = 1$.

If $\eta$ is a random variable with probability density
\begin{equation}
P(\eta) = \frac{1}{2}\left(  \delta (\eta - m - d)
	+ \delta(\eta - m + d)\right)
\label{eq:bernoulli}
\end{equation}
the solution of (\ref{eq:forr}) is straightforward:
\begin{equation}
r_{\rm opt} = \frac{1}{2} \left({ 1+\rho \over 1 + \rho -{\rm e}^{m+d} } +
{ 1+\rho \over 1 + \rho- {\rm e}^{m-d} }\right) \; .
\label{eq:rbernoulli}
\end{equation}
In more complicated cases we need to solve the equation (\ref{eq:forr})
numerically. However, for small mean and variance of $\eta$
approximative analytical formulae are fairly accurate
\cite{mas_zha_98,mas_zha_98a}. We found, that equally good
approximation is obtained, 
if we set $m=<\eta>$ and $d=\sqrt{<\eta^2>-<\eta>^2}$ in the
Eq. \ref{eq:rbernoulli}. 

 In the
next section we investigate the method with real data. Section
\ref{sec:transcosts} shows the influence of the transaction costs. In Sec.
\ref{sec:corr} a generalization of the method for the case of
time-correlated price is shown. Finally, in Sec. \ref{sec:concl}
we discuss the obtained results.

\section{Two-time optimal strategies}
\label{sec:twotime}

In the previous section we supposed the following procedure:
the investor takes the stock price data and extracts some statistical
information from them. This information is then plugged into theoretical
machinery, which returns the suggested number $r$. However, we may also
follow different path, which should be in principle equivalent, but
in practice it looks different.

Namely, suppose we observe the past evolution of the stock price
during some period  starting at time $t_1$ and finishing at time $t_2$ 
(most probably 
it will be the present moment, but not necessarily). Then, we imagine that
at time $t_1$ an investor started with capital $W_{t_1}=1$ and 
during that period followed the strategy determined by
certain value of $r$. We compute his/her capital $W_{t_2}(r)$ at final time
and find the maximum of  the final capital $W_{t_2}(r)$ with respect
to $r$. We call the  
value $r_{\rm opt}$ maximizing the final capital two-time optimal strategy.
Optimum strategy in the past can be then used as 
predicted optimal strategy for the future. 

\slaninafiginline{
\begin{figure}[hb]
  \centering
  \vspace*{\slaninafigspace}
  \includegraphics{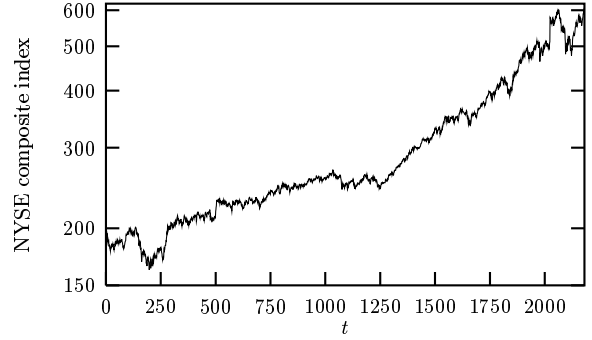}
  \caption{Time evolution of the NYSE composite index. Time is
           measured in working days from $t=0$ which is 
           2 January 1990 to $t=2181$, which is 
           31 December 1998. The vertical axis is in logarithmic 
           scale.}
  \label{fig:index}
\end{figure}
}
The capital at time $t_2$ is again
\begin{equation}
W_{t_2}(r) = \prod_{t=t_1}^{t_2-1}
(1+\rho + r ({\rm e}^{\eta_t}-1-\rho))\; 
\end{equation}
and its maximization with respect to $r$ leads to equation
\begin{equation}
g^\prime(r_{\rm opt}) = \sum_{t=t_1}^{t_2-1} 
{{\rm e}^{\eta_t}-1-\rho   \over 1+\rho + r_{\rm opt} ({\rm e}^{\eta_t}-1-\rho)} = 0
\end{equation}
which gives the optimal investment ratio $r_{\rm opt}(t_1,t_2)$ as a function of initial time $t_1$ and final time $t_2$. Note that it
is an analog of the equation (\ref{eq:forr}) but we deal with 
time averages here, not with sample averages as before.
This is also another justification of the procedure
of maximizing $<\log(W_{t+1}/W_t)>$ instead of $<W_{t+1}/W_t>$.

For comparison with reality we took the daily values of the New York 
Stock Exchange (NYSE) composite index. The time is measured in working
days. The period studied started on 2 January 1990
($t=0$) and finished on   
31 December 1998 ($t=2181$). The time evolution of the index $x(t)$ is
shown in Fig.  
\ref{fig:index}.
The values of $\eta$ are determined by $\exp(\eta_t)=x(t+1)/x(t)$. 

%From this data we calculated the time-dependent trend $m(t)$ and
%volatility $d(t)$  
%using exponentially weighted time averages \cite{ca_ma_zha_97}, which is
%%
%\begin{equation}
%m(t+1) = (1-\lambda)\ln(x(t+1)/x(t))+\lambda m(t)
%\end{equation}
%%
%for the trend and
%%
%\begin{eqnarray}
%m_2(t+1)&=&(1-\lambda)(\ln(x(t+1)/x(t)))^2+\lambda m_2(t)\\
%d(t+1) &=& \sqrt{m_2(t+1)-(m(t+1))^2}
%\end{eqnarray}
%%
%for the volatility. We used the parameter $\lambda=0.99$ which
% corresponds to effectively averaging over 100 days.
% The
%results are in fig.  \ref{fig:md}. We can see that the typical value of
% $d$ lies between 0.005 and 0.01.

The data of NYSE composite index were analyzed by calculating the
two-time optimal 
strategies $r_{\rm opt}(t_1,t_2)$.
As a typical example of the behavior observed, for  initial time $t_1=300$
we vary the final time $t_2$ up to 2180. We used the   
interest rate 6.5\% per 250 days (a realistic value for approximately
1 year). In this case we neglect the
transaction costs, $\gamma=0$. The influence of non-zero transaction
costs will be investigated in Sec. \ref{sec:transcosts}.
The results are in Fig. \ref{fig:rt-tax-loan}(a).
We investigated also the possibility that the investment ratio goes 
beyond the limits 0 and 1, which means that the investor borrows
either money or stock. We imposed the interest rate 8\% on the loans
and calculated again the optimal $r$. The results are in 
Fig. \ref{fig:rt-tax-loan}(c).
We can see several far-reaching excursions above 1 and some also below 0, 
which indicates that quite often the optimal strategy requires 
borrowing 
considerable amount of money or stock. 

\slaninafiginline{
\begin{figure}[hb]
  \centering
  \vspace*{\slaninafigspaceh}
  \includegraphics{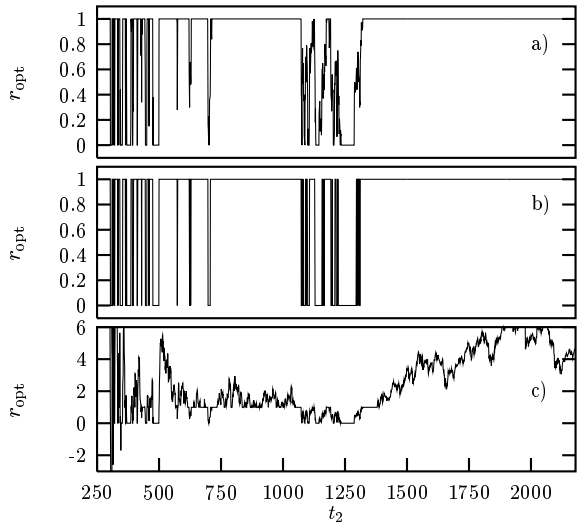}
  \caption{The two-time optimal investment ratio for interest rate 6.5\%
           per 250 days. The initial
           time is 300. The transaction costs are $\gamma=0$ (a) and
           $\gamma=0.005$ (b). In (c), loans are allowed with interest
rate 8\% per 250 days, transactions costs are $\gamma=0$.}
  \label{fig:rt-tax-loan}
\end{figure}}
An important conclusion may be drawn from the results obtained:
the optimal strategy $r_{\rm opt}(t_1,t_2)$ as a function of the final
time $t_2$ does not follow any smooth trajectory. On the contrary,
the dependence is extremely noisy, as can be seen very well in the
Fig. \ref{fig:rt-tax-loan}. Moreover, the strategy is very sensitive
to initial 
conditions. If we compare the strategy $r_{\rm opt}(t_1,t_2)$ and $r_{\rm
opt}(t_1+\Delta t,t_2)$ 
for slightly different initial time, big differences are found in
regions, where the strategy is non-trivial $(0<r_{\rm opt}<1)$. In
Fig. \ref{fig:differ} we show for $\Delta t=1$ the average difference
in optimal strategy 
\begin{equation}
\Delta r_{\rm opt}(t) = \left< |r_{\rm opt}(t_1,t_1+t) -r_{\rm
opt}(t_1+1,t_1+t)|\right>
\end{equation}
where the average is taken over all initial times $t_1$ with the
constraint, that we take into account only the points where both
optimal strategies $r_{\rm opt}(t_1,t_1+t)$ and $r_{\rm 
opt}(t_1+1,t_1+t)$ are non-trivial.
\slaninafiginline{
\begin{figure}[hb]
  \centering
  \vspace*{\slaninafigspace}
  \includegraphics{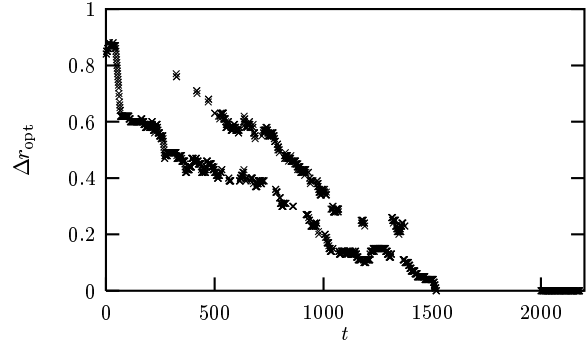}
  \caption{Average difference in optimal strategy when the initial
           times differ  by 1 day. Only points where the strategies
           are non-trivial are taken into account.}
  \label{fig:differ}
\end{figure}}

Due to poor statistics, the data are not very smooth. We can also
observe apparent two branches of the dependece, which is caused by
superimposing data from different portions of the time evolution of
the index. However, despite of the poor quality of the data, we 
can conclude, that even after a period as long as
1000 days (approximately 4 years) the difference of 1 day in the
starting time leads to difference in optimal strategy as large as
about 0.2. This finding challenges the reliability of the investment
strategy based on finding optimal investment ratio $r$.

Moreover, we can see that if loans are prohibited, 
there are long periods where the optimal strategy is trivial ($r_{\rm
opt} = 0$  
or $r_{\rm opt} =1 $). We investigated the whole history
of the NYSE composite index shown in Fig. \ref{fig:index} and determined,
for which pairs $(t_1,t_2)$ the optimal
strategy $r_{\rm opt}(t_1,t_2)$ is non-trivial.  
In the Fig. \ref{fig:tt} each dot represents such pair. (In fact, not every
point was checked: the grid $5\times 5$ was used, i.~e. only such 
times which are multiples of 5 were investigated.)

We can observe large empty regions, which indicate absence of
a non-trivial investment. In order to understand the origin of such
empty spaces, let us consider a simple model. Suppose we have the
random variable distributed according to (\ref{eq:bernoulli}),
%
%\begin{equation}
%P(\eta) = \frac{1}{2}(\delta(\eta-m-d) + \delta(\eta-m+d))
%\end{equation}
%
and $\rho = 0$.
Then the conditions for the existence of non-trivial
optimal strategy between $t_1=0$ and $t_2=N$ are
\begin{equation}
g^\prime(0) = \sum_{t=0}^{N-1} ({\rm e}^{\eta_t}-1) > 0
\end{equation}
and
\begin{equation}
g^\prime(1) = \sum_{t=0}^{N-1} (1-{\rm e}^{-\eta_t}) < 0\; .
\end{equation}
\slaninafiginline{
\begin{figure}[hb]
  \centering
  \vspace*{\slaninafigspace}
  \includegraphics{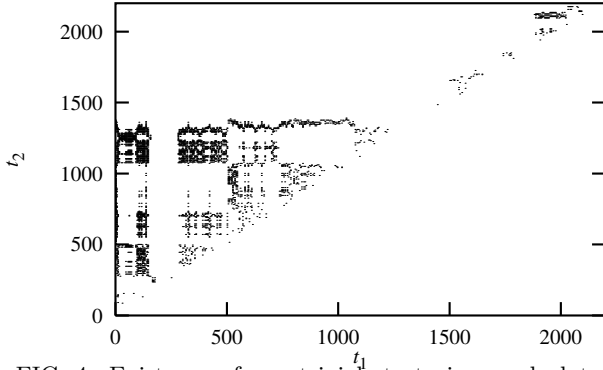}
  \caption{Existence of non-trivial strategies: each dot represents
           starting and final time between which a non-trivial optimal strategy
           is found.}
  \label{fig:tt}
\end{figure}}

Let us compute the probability $p_{\rm nt}$ that both of these conditions
are satisfied.
We have 
\begin{equation}
g^\prime(0) = N({\rm e}^m\cosh d -1) + 
{\rm e}^m\sinh d \sum_{t=0}^{N-1} z_t
\end{equation}
and
\begin{equation}
g^\prime(1) = N(1-{\rm e}^{-m}\cosh d ) +
 {\rm e}^{-m}\sinh d \sum_{t=0}^{N-1} z_t
\end{equation}
where $z$'s can have values +1 or -1 with probability 1/2.
The sum $\sum_{t=0}^{N-1} z_t  $ has binomial distribution, 
and for large $N$
we can write
\begin{equation}
p_{\rm nt} = \int_{
-\sqrt{N} (\coth d - {\rm e}^{-m}/\sinh d)
}^{
\sqrt{N} (\coth d - {\rm e}^{m}/\sinh d)
}
{{\rm d}\zeta \over\sqrt{2\pi}}\;\exp(-{\zeta^2\over 2})\;\; .
\end{equation}

We can see immediately that $p_{\rm nt}$ has a value close to 1 for
the number of time steps at least
\begin{equation}
N\simeq d^{-2}\;\; .
\end{equation}
For the data in Fig. \ref{fig:index} we found 
$d\simeq 0.01$, which 
means $N\simeq 10000$ days, or 40 years. This is thus an estimate
of how long we need to observe the stock price before a reliable
strategy can be fixed. However, during such a long period the market 
changes substantially many times. That is why no simple strategy
of the kind investigated here
can lead to sure profit.

\section{Transaction costs}
\label{sec:transcosts}

We investigated the influence of the transaction costs $\gamma$ and
time lag $M$ between transactions.  We found nearly no dependence on
$M$, but the dependence on $\gamma$ is rather strong. It can be
qualitatively seen in Fig. \ref{fig:rt-tax-loan}(b). When we compare the
optimal strategy for $\gamma=0$ and $\gamma=0.005$, we can see, that
already transaction costs $0.5\%$ decrease substantially the fraction
of time when the strategy is non-trivial. We investigated the
dependence of the fraction $f_{\rm nontrivial}$ of
time pairs $(t_1,t_2)$ between which a nontrivial strategy exists on the
transaction costs. We have found
that it decreases with
$\gamma$ and reaches negligible value for $\gamma \ge 0.006$. This
behavior is shown in Fig. \ref{fig:nontrivial}.

\slaninafiginline{
\begin{figure}[hb]
  \centering
  \vspace*{\slaninafigspace}
  \includegraphics{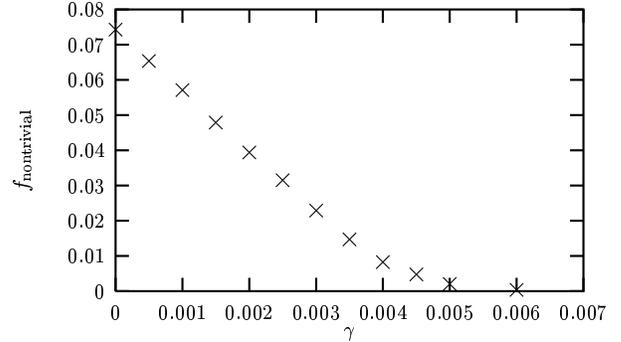}
  \caption{The dependence of the fraction of time pairs, between which
           a non-trivial  investment optimal strategy exists, on the
           transaction cost.  The time interval investigated is from
           time 0 to time 1600. }
  \label{fig:nontrivial}
\end{figure}}
 The
explanation of this behavior lies in the fact, that the transaction
costs introduce 
some friction in the market, which means that large changes of the
investment ratio are suppressed. Because the investment ration is
mostly 0 or 1 even for $\gamma=0$, this implies that changing $r$ from
0 or 1 to a non-trivial value is even harder for $\gamma > 0$ and a
non-trivial investment becomes nearly impossible for large transaction
costs.

\section{Investment in presence of correlations}
\label{sec:corr}

In order to improve the strategy based only on the knowledge
of the distribution of $\eta$, we would like to
investigate a possible profit taken from the short-time correlations.

Imagine again the simplest case, when $\eta$ can have only two values,
$\eta^+ = m+d$ and $\eta^- = m-d$. However, now $\eta_t$ and $\eta_{t-1}$ may be 
correlated and we suppose the following probability distribution 
$P(\eta_{t-1},\eta_t) = 1/4 + c$ if $\eta_{t-1} = \eta_t$ and
$P(\eta_{t-1},\eta_t) = 1/4 - c$ if $\eta_{t-1} \neq \eta_t$.
The parameter $c$ quantifies the short-time correlations.

At time $t$ the strategy $r(\eta_{t-1})$ 
should depend on the value of $\eta$ in the previous step. In our 
simplified situation we have only two possibilities, $r^+ = r(\eta^+) $
and $r^- = r(\eta^-)$. The problem then reduces to maximization
of the typical gain
\begin{equation}
g(r^+,r^-) = <\ln (1+\rho + r(\eta_{t-1})({\rm e}^{\eta_t}-1-\rho))>
\end{equation}
which leads to decoupled equations for $r^+_{\rm opt}$ and $r^-_{\rm opt}$
\begin{eqnarray}
{\partial g (r^+_{\rm opt},r^-_{\rm opt})\over \partial r^+} =&
(\frac{1}{4}+c) { {\rm e}^{m+d} - 1 - \rho    \over 
1 + \rho + r^+_{\rm opt} ({\rm e}^{m+d} - 1 - \rho )}+\\
 &(\frac{1}{4}-c) { {\rm e}^{m-d} - 1 - \rho    \over 
1 + \rho + r^+_{\rm opt} ({\rm e}^{m-d} - 1 - \rho )} = 0
\end{eqnarray}
\begin{eqnarray}
{\partial g (r^+_{\rm opt},r^-_{\rm opt})\over \partial r^-} = &
(\frac{1}{4}+c) { {\rm e}^{m-d} - 1 - \rho    \over 
1 + \rho + r^-_{\rm opt} ({\rm e}^{m-d} - 1 - \rho )}
+\\
&(\frac{1}{4}-c) { {\rm e}^{m+d} - 1 - \rho    \over 
1 + \rho + r^-_{\rm opt} ({\rm e}^{m+d} - 1 - \rho )} = 0 \;\; .
\end{eqnarray}
The solution is a straightforward generalization of
Eq. (\ref{eq:bernoulli}). 

The above procedure works equally well even in the case of more
complicated 
time correlations. For example we may imagine that the price evolution
is positively correlated over two time steps, i.~e.  
${\rm Prob}(\eta_{t-2} = \eta_t) > 1/2$, while 
${\rm Prob}(\eta_{t-1} = \eta_t) = 1/2$. Generally, we have some joint 
probability distribution for the past and present $P(\eta^<,\eta)$, where
we denote $\eta^< =[...,\eta_{t-3},\eta_{t-2},\eta_{t-1}]$ and
$\eta=\eta_t$. 
The typical gain becomes a functional depending
on the strategy $r(\eta^<)$ which itself depends on the past price history.

However, maximizing this functional by looking for its stationary point
leads to very simple set of decoupled equations for the strategies
\begin{equation}
\int {\rm d}\eta\; P(\eta^<,\eta){ {\rm e}^\eta - 1 - \rho    \over 
1 + \rho + r_{\rm opt}(\eta^<) ({\rm e}^\eta - 1 - \rho )} = 0\; .
\end{equation}

In the simplest case, when we assume that the strategy depends only on the
sign of $\eta$ in the previous step, we performed the analysis on the
NYSE composite index shown in the Fig. \ref{fig:index}. We found
optimal pairs $[r^+_{\rm opt},r^-_{\rm opt}]$. Contrary to the case
when correlations were not taken into account, no non-trivial
investment strategy was found. So, instead to improve the 
method of Ref. \cite{ma_mas_zha_98}, the 
applicability of this method is further discredited.

\section{Conclusions}
\label{sec:concl}

We investigated the method of finding the optimal investment strategy
based on the Kelly criterion. We checked the method on real data based
on the time evolution of 
the New York Stock Exchange composite index. We found, that it is
rarely possible to find an
optimal strategy which would be stable at least for a short period of
time. There are several reasons, which discredit the method based on
the Kelly criterion. First, the optimal investment ratio fluctuates
very rapidly in time. Second, it depends strongly on the time, when
the investment strategy started to be applied. The difference of 1 day
in the starting moment makes substantial difference even after 1000
days of investment. Third, the fraction of days, for which a
non-trivial investment strategy is possible, is very low. This
fraction also decreases with transaction costs and reaches negligible
values for transaction costs about $0.6\%$. Taking into account
possible correlations in the time evolution of the index makes the
situation even less favorable, reducing further the fraction of times, when a
non-trivial investment is possible.

We conclude, that straightforward application of the investment
strategy based on the Kelly criterion would be very difficult in real
conditions. The question remains, whether there are other optimization
schemes, which would lead to more certain investment strategies. It
would be also interesting to apply the approach used in this paper in
order to check the reliability of the option-pricing strategies.

\acknowledgments{I am obliged to Y.-C. Zhang, M. Serva, and M. Kotrla for many
discussions and stimulating comments. I wish to thank to the Institut
de Physique Th\'eorique, Universit\'e de Fribourg, Switzerland, for
financial support.
}


\begin{thebibliography}{99}
\bibitem{mantegna_91}
R. N. Mantegna,
 Physica A
{\bf 179}, 232
(1991).

\bibitem{am_bu_ha_97}
L.~A.~N.~Amaral, S.~V.~Buldyrev, S.~Havlin, P.~Maas, M.~A.~Salinger, H.~E.~Stanley, and M.~H.~R.~Stanley,
 Physica A
{\bf 244}, 1
(1997).

\bibitem{li_ci_me_pe_sta_97}
Y.~Liu, P.~Cizeau, M.~Meyer, C.-K.~Peng, and H.~E.~Stanley,
 Physica A
{\bf 245}, 437
(1997).

\bibitem{go_me_am_sta_98}
P. Gopikrishnan, M. Meyer, L. A. N. Amaral, and H. E. Stanley,
 Eur. Phys. J. B
{\bf 3}, 139
(1998).

\bibitem{ga_ca_ma_zha_97}
S.~Galluccio, G.~Caldarelli, M.~Marsili, and Y.-C.~Zhang,
 Physica A
{\bf 245}, 423
(1997).

\bibitem{va_bo_mi_au_98}
N. Vandewalle, P. Boveroux, A. Minguet, and M. Ausloos,
 Physica A
{\bf 255}, 201
(1998).

\bibitem{so_jo_97}
D.~Sornette and A.~Johansen,
 Physica A
{\bf 245}, 411
(1997).

\bibitem{sornette_98d}
D. Sornette,
 Eur. Phys. J. B
{\bf 3}, 125
(1998).

\bibitem{bou_so_94}
J.-P.~Bouchaud and D.~Sornette,
 J. Phys. I France
{\bf 4}, 863
(1994).

\bibitem{bou_po_97}
J.-P.~Bouchaud and M.~Potters,
 Th\'eorie des risques financiers
(Al\'ea, Saclay, 1997).

\bibitem{ga_bou_po_98}
S. Galluccio, J.-P. Bouchaud, and M. Potters,
 Physica A
{\bf 259}, 449
(1998).

\bibitem{bouchaud_98}
J.-P. Bouchaud,
 cond-mat/9806101.

\bibitem{cont_98}
R. Cont,
 cond-mat/9808262, {\rm to appear in: }  Econophysics: an emerging science, eds. J. Kert\'esz and I. Kondor 
(Kluwer Academic Publishers, Dordrecht, 1998).

\bibitem{ca_ma_zha_97}
G.~Caldarelli, M.~Marsili, and Y.-C.~Zhang,
 Europhys. Lett.
{\bf 40}, 479
(1997).

\bibitem{ba_pa_shu_97}
P. Bak, M. Paczuski, and M. Shubik,
 Physica A
{\bf 246}, 430
(1997).

\bibitem{cha_zha_97}
D.~Challet and Y.-C. Zhang,
 Physica A
{\bf 246}, 407
(1997).

\bibitem{sa_ta_98}
A.-H. Sato and H. Takayasu,
 Physica A
{\bf 250}, 231
(1998).

\bibitem{co_bou_97}
R.~Cont and J.-P.~Bouchaud,
 cond-mat/9712318.

\bibitem{bou_co_98}
J.-P. Bouchaud and R. Cont,
 Eur. Phys. J. B 
{\bf 6}, 543
(1998).

\bibitem{ma_mas_zha_98}
M. Marsili, S. Maslov, and Y.-C. Zhang,
 Physica A
{\bf 253}, 403
(1998).

\bibitem{kelly_56}
J.~L.~Kelly,
 The Bell System Technical Journal
{\bf 35}, 917
(1956).

\bibitem{thorp_71}
E.~O.~Thorp,
 {\rm in: } American Statistical Association, Bussines and Economics Statistics section, Proceedings 
(1971) p. 599.

\bibitem{mas_zha_98}
S. Maslov and Y.-C. Zhang,
 Int. J. Theor. Appl. Finance
{\bf 1}, 377
(1998).

\bibitem{mas_zha_98a}
S. Maslov and Y.-C. Zhang,
 Physica A
{\bf 262}, 232
(1999).

\bibitem{ba_pa_se_vu_98}
R. Baviera, M. Pasquini, M. Serva, and A. Vulpiani,
 cond-mat/9804297.

\bibitem{au_ba_ha_se_vu_98}
E. Aurell, R. Baviera, O. Hammarlid, M. Serva, and A. Vulpiani,
 cond-mat/9810257.

\bibitem{serva_98}
M. Serva,
 cond-mat/9810091, {\rm submitted to } Int. J. Theor. Appl. Finance.

\bibitem{deutsch_94}
J. M. Deutsch,
 Physica A
{\bf 208}, 433
(1994).

\bibitem{sornette_98c}
D. Sornette,
 Phys. Rev. E
{\bf 57}, 4811
(1998).

\bibitem{so_co_97}
D.~Sornette and R.~Cont,
 J. Phys I France
{\bf 7}, 431
(1997).

\bibitem{sornette_97a}
D.~Sornette,
 Physica A
{\bf 250}, 295
(1998).

\end{thebibliography}
\end{document}